\documentstyle[cite]{EuroPhys}

\def\etal{{\hbox{{\tenit\ et al.\/}\tenrm :\ }}}

\def\stars{\bigskip\centerline{***}\medskip}

\newif\ifboo \boofalse

\def\Review#1{\boofalse{\it #1},}
\def\Name#1{{\sc #1},}
\def\Vol#1{\ifboo Vol. {\bf #1}\else{\bf #1}\fi}
\def\Year#1{\ifboo #1\else(#1)\fi}
\def\Book#1{\bootrue{\it #1},}
\def\Page#1{\ifboo {\rm p. #1}\else{\rm #1}\fi}

\begin{document}

\def\viipt{}
\def\viiipt{}
\def\ixpt{}
\def\tenrm{}
\def\tenit{}
\input epsf
\epsfverbosetrue

\def\livo{LiVO$_2$ }
\def\natio{NaTiO$_2$ }

\euro{ }{ }{ }{ }
\Date{ }
\shorttitle{S. EZHOV \etal ORBITAL POLARIZATION IN LiVO$_2$ AND NaTiO$_2$}

\title{Orbital polarization in \livo and \natio}
\author{S.Yu.Ezhov\inst{1}, V.I.Anisimov\inst{1},
        H.F.Pen\inst{2}, D.I.Khomskii\inst{2}, G.A.Sawatzky\inst{2}}
\institute{
      \inst{1}Institute of Metal Physics, GSP-170, Ekaterinburg, Russia\\
      \inst{2}Laboratory of Applied and Solid State Physics,
              Materials~Science~Centre, University~of~Groningen, 
              Nijenborgh~4, 9747~AG~Groningen, The~Netherlands
           }
\rec{}{}
\pacs{
\Pacs{71}{15Mb}{Density functional theory, local density approximation}
\Pacs{71}{27$+$a}{Strongly correlated electron systems; heavy fermions}
\Pacs{71}{20$-$b}{Electron density of states and band structure of crystalline solids}
     }
\maketitle

\begin{abstract}
We present a band structure study of orbital polarization and
ordering in the two-dimensional triangular lattice transition metal
compounds   \livo and \natio. It is found that while in
\natio the degeneracy of $t_{2g}$ orbitals is lifted due to the trigonal
symmetry of the crystal and the strong on cite Coulomb interaction, 
in \livo orbital degeneracy remains and orbital ordering
corresponding to the trimerization of the two-dimensional lattice develops.
\end{abstract}

It is well known that transition metal compounds with orbital
degeneracy will in some way restructure to remove that orbital
degeneracy in the ground state. Well known is the example of a
two-fold orbitally degenerate case of divalent Cu in octahedral
symmetry with one hole in a $e_g$-like orbital. A similar case is
trivalent Mn in $O_h$ symmetry as in the now well known collossal
magnetoresistance materials. In these so called strong Jan
Teller systems local lattice distortions determine the
type of orbital ordering. It is also well established that
the relative spatial orientation of occupied orbitals on
neighboring ions determines not only the magnitude but also the
sign of the exchange interactions governing the magnetic
structure of the system\cite{kugel82}. In the
early $3d$ transition metal compounds only the $t_{2g}$ orbitals are
occupied leaving us with three-fold degeneracy in the cases of Ti$^{3+}$
and V$^{3+}$  with one and two $3d$ electrons respectively assuming
also $O_h$ symmetry. In contrast to the $e_g$ orbitals the bonding to
the neighboring O~$2p$ orbitals is much weaker, bandwidths are
much smaller and therefore the removal of the orbital
degeneracy may be more subtle. It has for example recently
been suggested that the orbital degeneracy in LiVO$_2$ can be
lifted by a particular kind of orbital ordering driven by the
nearest neighbor exchange interactions\cite{livo2}. The orbital ordering
proposed there is one which simultaniously removes the
frustration in the spin Hamiltonian of this triangular two-dimensional
lattice and results in a non magnetic singlet ground
state. Another much discussed two-dimensional
triangular lattice spin system is NaTiO2 with spin $\frac{1}{2}$ per Ti ion,
which in case of uniform Heisenberg interaction would be a frustrated 
magnetic system, making it one of the few remaining possible
examples of a resonating valence bond (RVB) ground state\cite{anders}. However also
here the orbital degeneracy could have been lifted by orbital ordering
resulting in a non-uniform Heisenberg spin Hamiltonian and a
subsequent removal of the frustration. In this paper we present
a detailed band structure study of these two systems both with
and without the inclusion of the on site $d$-$d$ Coulomb interaction
in a so called LDA+U approximation.  We find that although
LiVO$_2$ behaves as suggested by Pen~{\it et~al}\cite{livo2} NaTiO$_2$ should 
not be considered as orbitally degenerate and remains a candidate for an
RVB ground state\cite{anders}. 

\begin{figure}
\epsfxsize=12.0truecm\centerline{\epsfbox{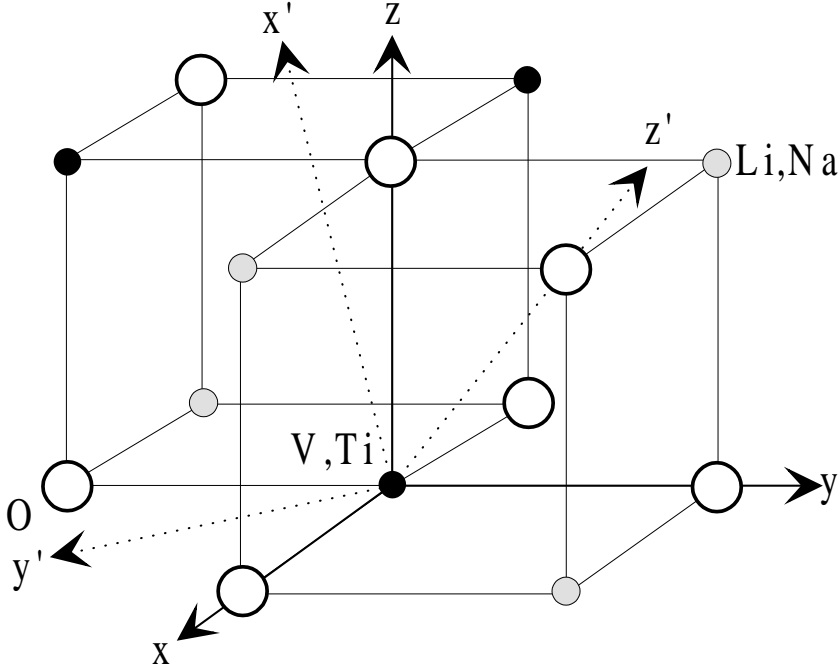}}
\caption{The fragment of the crystal structure of $NaTiO_2$ and
$LiVO_2$. $x,y,z$ are along the cubic crystal axes, the $z'$ axis is the
trigonal threefold rotation axis.}
\label{fig:crystal}
\end{figure}

LiVO$_2$ and NaTiO$_2$ crystallize in an ordered rock salt-like structure
with alternating [111] planes of Li(Na),O,V(Ti),O ions 
as shown in fig.~\ref{fig:crystal}. Each of these layers forms a triangular two 
dimensional lattice and since the layers containing the magnetic ions are so
well seperated  the exchange interactions between layers is very
small compared to that within a layer. Peculiar in these compounds is a change 
from a high temperature paramagnetic phase and a Curie-Weiss susceptibility
with a large negative effective Curie temperature (which corresponds
to strong antiferromagnetic coupling between local moments) to a low
temperature nonmagnetic system without any sign of long-range
order\cite{bongers57,exper}.  This is usually explained by the
frustration of the triangular lattice antiferromagnet leading to the 
possibility of an RVB ground state with a quantum liquid 
of randomly distributed spin singlet pairs\cite{anders}.  However
such models do not take into account orbital degeneracy and the possibility of
orbital ordering, which can qualitatively change the type of
exchange interactions as discussed above.

While the nearest neighbors of the transition metal ion (oxygen
atoms) are arranged in an almost perfect octahedron with $O_h$
symmetry, the overall symmetry of the crystal structure is trigonal, and in
that case the $t_{2g}$ level is split into a nondegenerate $A_{1g}$
and double degenerate $E_g$ levels. (Note: in further discussion
$e_g$ and $E_g$ have different meanings). The $t_{2g}$-orbitals can be
represented as the set $(d_{xy},d_{yz},d_{zx})$ in the coordinate system
with axes pointing towards the O-neighbors (fig.~\ref{fig:crystal}). In this
coordinate system the $A_{1g}$ and $E_g$-orbitals have the following 
form\cite{terakura}: 
\begin{eqnarray}
\label{eq:a1-e}\begin{tabular}{cccll}
$A_{1g}:(d_{xy}+d_{yz}+d_{zx})/\sqrt{3}$ &\makebox[1cm]{} 
&$E_g:$ &
$(1/\sqrt{2)}(d_{zx}-d_{yz})$ \\
 &  &  & $(1/\sqrt{6})(-2d_{xy}+d_{yz}+d_{zx})$\\
\end{tabular}
\end{eqnarray}
\begin{figure}
\epsfxsize=12.0truecm\centerline{\epsfbox{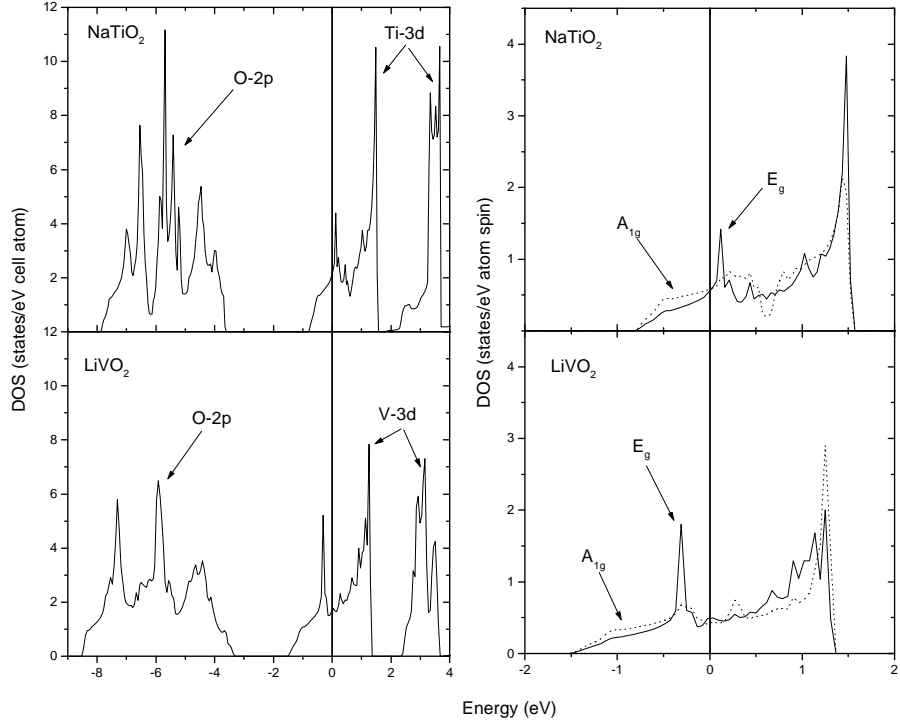}}
\caption{Left panel -- the total densities of states from LDA calculations.
Right panel -- the partial densities of states from LDA calculations: dotted line -- A$_{1g}$,
solid line -- E$_g$. The Fermi energy is the zero energy. }
\label{fig:lda}
\end{figure}
The resulting orbital order will
strongly depend on how this trigonal splitting will affect the
resulting orbital polarization of the partially filled $t_{2g}$
band. If the centre of gravity of the $A_{1g}$ band is lower than
$E_g$, then in the case of the $d^1$ configuration the degeneracy will
be lifted, but the $d^2$ configuration will be still degenerate provided
that the Hunds rule exchange is larger than this splitting
which is certainly expected to be the case.  If the $E_g$ level is lower 
than $A_{1g}$, then the $d^2$
configuration in the high spin  state will be nondegenerate.

In order to look into this problem we have performed band
structure calculations for LiVO$_2$ and NaTiO$_2$ by LMTO method\cite{lmto} 
in standard Local Density Approximation (LDA)\cite{lda} and 
also in LDA+U approach\cite{lda+u} which allows to take into account
Coulomb correlations between $3d$ electrons of transition metal ions.
We used the crystal structures and atomic positions as given by\cite{strucV,strucTi}.
The total DOS obtained in LDA and presented in the left panel of fig.~\ref{fig:lda}, shows 
the O $2p$ band between -8 and -4
eV, (Ti,V) $3d$ band of $t_{2g}$ symmetry crossing the Fermi level  between
-1 and 1 eV, and of $e_g$ symmetry around 3 eV well above the Fermi energy. 
As one can see, indeed the $t_{2g}$-$e_g$ crystal field splitting is larger
than the bandwidth and V $3d$ subbands of the $t_{2g}$ and $e_g$ symmetry are 
separated by a gap. We should mention
that the $e_g$ band width is even smaller than that of $t_{2g}$,
this is because the $e_g$-$e_g$ hopping of the neighboring tranisition metal 
ions goes through oxygen atoms and the angle of Me-O-Me bond is 
nearly 90$^\circ$, so the $e_g$-$e_g$ hybridization is small. 
In spite of the fact that there is a trigonal distortion
of the lattice the $t_{2g}$ band is not splitted due to the
trigonal symmetry into $A_{1g}$ and $E_g$ subbands and a  more delicate
quantitative analysis is needed to clarify this problem.

In the right panel of fig.~\ref{fig:lda} the partial DOS for decomposition of the
$t_{2g}$ band into orbitals of the $A_{1g}$ and $E_g$ symmetry
are presented.  One can see that the situation can not be described
in the simple terms of $A_{1g}$-$E_g$ "splitting":  both curves have
the same width and they are approximately in the same energy
region. We can estimate the actual splitting of the $A_{1g}$ and $E_g$ levels
by calculating the values of the centres of gravity of these bands.
\begin{figure}
\epsfxsize=12.0truecm\centerline{\epsfbox{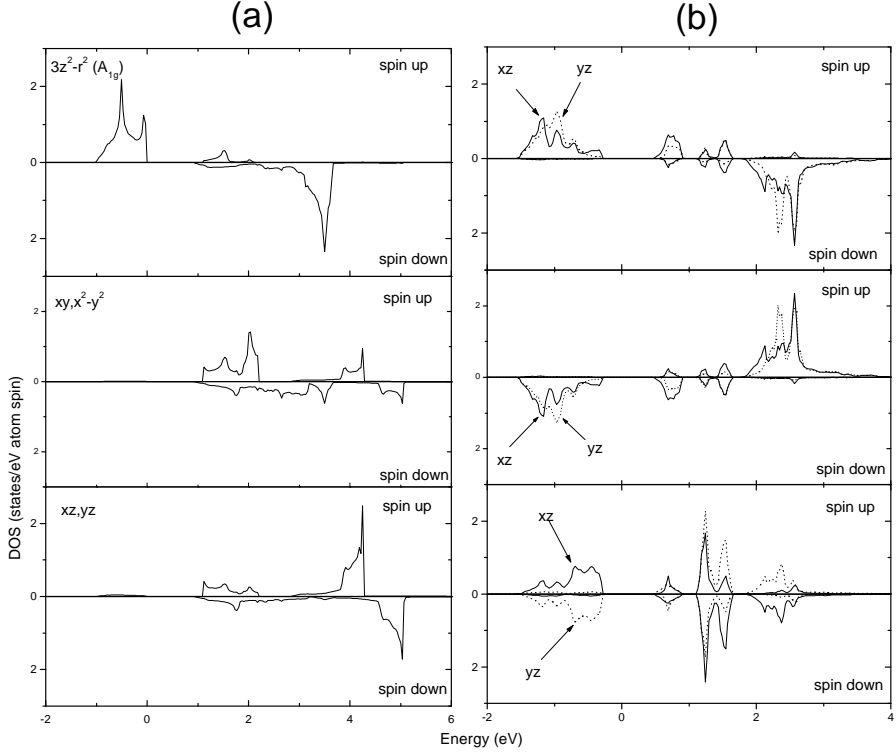}}
\caption{(a) -- the Ti($3d$) partial densities of states for
rhombohedral NaTiO$_2$ obtained from LDA+U calculations.  
(b) -- The V(3d) partial densities of states for LiVO$_2$ from LDA+U calculations.
The upper panel corresponds to the V atom with total
spin direction "up", the middle one for the V atom with "spin-down",
and the lower one to the V atom with zero moment.  Only xz and yz
orbitals in local coordinate system for each V atom
of the t$_{2g}$ subband are depicted.  On every V-atom $z$ axis
points to the oxygen atom above to the center of the V-triangle.  The
Fermi energy is the zero energy.}
\label{fig:ldau}
\end{figure}

In the case of NaTiO$_2$ with one $d$  electron the center of
gravity of the $A_{1g}$ band is 0.1~eV lower than that of the $E_g$ band. As a
result the occupied part of $t_{2g}$ band has slightly more $A_{1g}$
character than $E_g$, and the occupancy of orbitals are .25 and .20 per 
spin-orbital for $A_{1g}$ and for each $E_g$ correspondingly.  This means 
that the degeneracy of the $t_{2g}$ orbitals is essentially lifted but
the splitting is still much less than the band widths. This
small splitting is non the less important since as we will see
below if we turn on the $d$-$d$ Coulomb interaction in LDA+U the
$A_{1g}$ band will be occupied and the $E_g$ unocupied now with
a splitting mainly due to U. However the choice as to which band
is occupied and which one not is dictated by the small crystal
field splitting. The above result would also indicate that the
$A_{1g}$-$E_g$ local excitation energy would be only 0.1 eV or
so and would contribute to charge conserving excitonic-like
excitations. In this LDA+U calculation the $d$-$d$ Coulomb
interaction was found to be 3.6 eV (taking 
into account the screening of $t_{2g}$ electrons by $e_g$
electrons\cite{Ucalc}) which is much larger than the $t_{2g}$ band width
and leads to the localization of a single $d$ electron in the $A_{1g}$ orbital. 
\begin{figure}
\epsfxsize=12.0truecm\centerline{\epsfbox{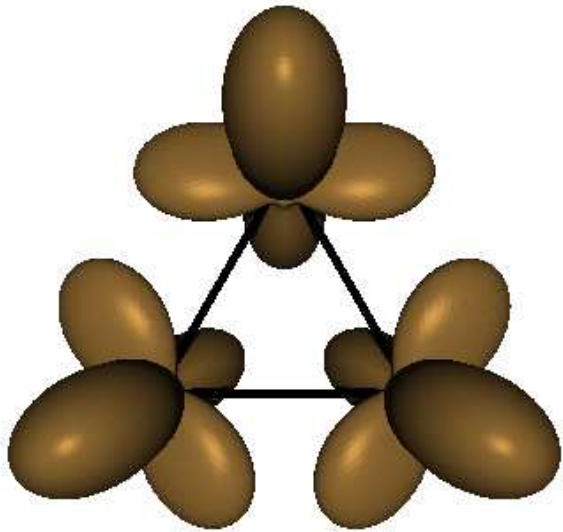}}
\caption{The t$_{2g}$ holes from LDA+U calculations for LiVO$_2$.
Only V atoms for one triangle in a hexagonal plane are drawn. The
view is from the point directly above V-triangle}
\label{fig:orbital}
\end{figure}

The LDA+U calculations were carried out for both antiferromagnetic and
ferromagnetic cases.  For the AFM case we choose the simplest
magnetic order with four nearest neighbors out of the six in the basal
plane having anti-parallel spin orientation and other two parallel.
Independent of the spin ordering a single 
$d$ electron in the $t_{2g}$ shell of the Ti ion turned out to be localized in the 
$A_{1g}$ orbital. The occupation numbers for the majority spin are 0.9 for 
$A_{1g}$ and 0.1 for $E_g$ for both FM and AFM cases. 
The Ti($3d$) projected density-of-states obtained from LDA+U
calculations is shown in fig.~\ref{fig:ldau}(a). 
So we can say from the fig.~\ref{fig:ldau}(a) 
that the LDA+U solution for NaTiO$_2$ is almost fully orbitally polarized.  
One can see from eq.~\ref{eq:a1-e} that the $A_{1g}$ orbital 
(d$_{3z^2-r^2}$ in fig.~\ref{fig:ldau}(a))
is symmetric in the hexagonal Ti-Ti plane and
the occupation of this orbital leads to the isotropic exchange. This
indicates that NaTiO$_2$ would still behave like a frustrated
spin system.

In the case of LiVO$_2$ the centre of gravity of the $A_{1g}$
band is only 0.025~eV 
lower then the centre of gravity of the $E_g$ band, and the
resulting occupancies are 0.37 and 0.36 for $A_{1g}$ and for each
$E_g$-orbitals correspondingly.  In this situation orbital degeneracy is
not lifted, since we now have two electrons one in a A$_{1g}$ and one in a
E$_g$ orbital, because the Hunds rule
exchange strongly favours the high spin state. As a result the 
appearance of some kind of orbital order can
be expected.  In\cite{livo2,Good1,Good2} the
formation of local spin singlets on trimers containing V-atom
triangles was suggested as the model explaining the low-temperature
nonmagnetic behavior of LiVO$_2$.  Those spin singlets were
stabilized by a  specific orbital order\cite{livo2}.

The LDA+U method is based on a mean-field approximation and
can not fully reproduce the essentially many-electron singlet wave
function, especially the correct energy difference of singlet-triplet
configurations.  However a single Slater determinant trial wave
function can still describe the basic relationship between spin and
orbital degrees of freedom.  To imitate trimer spin singlets we performed 
LDA+U calculations with spin-order of the type "up-down-zero" on every
triangle (closed circles on the fig.~\ref{fig:crystal}).
The self-consistent calculations resulted in the orbital
order of the same type as proposed in\cite{livo2} from model
calculations~[fig.~\ref{fig:ldau}(b)]:  on every V atom the occupied
orbitals are $xz$ and $yz$ if in a local coordinate system $z$
axis is directed towards the oxygen atom sitting just above the center
of V-triangle, and x and y axes are directed towards other oxygens
of an octahedron(fig.~\ref{fig:crystal}). We should also mention
the fact that the LDA+U calculations give the correct semiconducting
state for LiVO$_2$ [fig.~\ref{fig:ldau}(b)] instead of a metallic  state 
from "normal" LDA (left panel of fig.~\ref{fig:lda}).

In fig.~\ref{fig:orbital} the angular distribution of the
t$_{2g}$ hole is presented as was obtained from the LDA+U
calculations.  It indicates the same orbital order proposed
in\cite{livo2} (fig.~1(a) in\cite{livo2}): $xz$ and $yz$ orbitals are
occupied, the $t_{2g}$ hole is in $xy$ orbital in a local
coordinate system of every V atom.

Both LiVO$_2$ and NaTiO$_2$ were regarded as candidates for
realization of Anderson's "resonating valence bond" systems with
a quantum liquid of randomly distributed spin singlet pairs.  Our
results show that while in LiVO$_2$ more complicated trimer spin
singlets with corresponding orbital order are formed, no orbital
order due to the crystal field lifting of the orbital degeneracy is
present in \natio, and its magnetic properties are most
probably explained by a nondegenerate model, so that it is
indeed a good candidate for Anderson's RVB state.  What then is the
nature of the structural phase transition observed in \natio at
T$_c$=250, remains an open question.

Summarizing, we have shown that the degeneracy of the
$t_{2g}$-orbitals in \natio is lifted because of the trigonal symmetry of
the crystal and the large $d$-$d$ Coulomb interaction and no orbital
ordering occurs. In \livo orbital degeneracy 
remains in spite of the same trigonal distortion
as in \natio, and in effect the orbital ordering consistent with the
trimerization of the two-dimensional lattice takes place.

\stars
We thank Dr. S.J.Clarke for providing us with the detailed data of the
\natio crystal structure prior to publication.
This investigation was supported by the Russian Foundation
for Fundamental Investigations (RFFI grant 96 02-16167) and by
the Netherlands Organization for Fundamental Research on Matter
(FOM), with financial support by the Netherlands Organization for the
advance of Pure Science (NWO).


\vskip-12pt
\begin{thebibliography}{99}
%
\bibitem{kugel82}
\Name{K.I. Kugel and D.I. Khomskii}
\Review{Sov. Phys. Usp.} \Vol{25} \Page{231} \Year{1982}.  
%
\bibitem{livo2}  
\Name{H.F. Pen, J. van den Brink, D.I. Khomskii, G.A. Sawatzky} 
\Review{Phys. Rev. Lett.} \Vol{78} \Page{1323} \Year{1997}.
%
\bibitem{anders} 
\Name{P.W. Anderson} \Review{Science}
\Vol{235} \Page{1196} \Year{1987};
\Name{P.W. Anderson} \Review{Mater. Res. Bull.} 
\Vol{8} \Page{153} \Year{1973}.
%
\bibitem{bongers57} 
\Name{P.F. Bongers} \Book{Ph.D. thesis}
(University of Leiden)  \Year{1957}. 
%
\bibitem{exper} 
\Name{K. Kobayashi, K. Kosuge, S. Kashi}
\Review{ Mater. Res. Bull.}  \Vol{4} \Page{95} \Year{1969}; 
\Name{ L.P. Cardoso, D.E. Cox, T.A. Hewston, B.L. Chamberland} 
\Review{J.Solid State Chem.} \Vol{72} \Page{234} \Year{1988}.
%
\bibitem{terakura} 
\Name{K.Terakura, T.Oguchi, A.R.Williams, J.K\"ubler}
\Review{Phys. Rev. B} \Vol{30}  \Page{4734} \Year{1984}.
%
\bibitem{lmto} 
\Name{Andersen O.K.} \Review{Phys. Rev. B} \Vol{12}  \Page{3060} \Year{1975}.
%
\bibitem{lda} 
\Name{P. Hoenberg, W. Kohn} \Review{Phys. Rev.}
\Vol{ 136} \Page{B864} \Year{1964}; 
\Name{ W. Kohn, L.J. Sham} \Review{Phys. Rev.}
\Vol{140}  \Page{A1133} \Year{1965}.
%
\bibitem{lda+u} 
\Name{V.I. Anisimov, J. Zaanen, Andersen O.K.}  
\Review{Phys. Rev. B} \Vol{44} \Page{943} \Year{1991}; 
\Name{V.I. Anisimov, F. Aryasetiawan, A.I. Lichtenstein}
\Review{J. Phys.: Condens. Matter} \Vol{9} \Page{767} \Year{1997}.
%
\bibitem{strucV}
\Name{Katsushiro Imai, Hiroshi Sawa, Masayoshi Koike,
Masashi Hasegawa, Humihiko Takei} 
\Review{J. Solid State Chem.} \Vol{114} \Page{184} \Year{1995}.
%
\bibitem{strucTi} 
\Name{S.J. Clarke, A.C. Duggan, A.J. Fowkes, A.
Harrison, R.M. Ibberson, M.J. Rosseinsky}
\Review{Chem. Comm.} \Year{1996} \Page{409};
\Name{S.J.Clarke}  (private communication).
%
\bibitem{Ucalc} 
\Name{V.I. Anisimov, O.Gunnarsson} 
\Review{Phys. Rev. B} \Vol{43} \Page{7570} \Year{1991}; 
\Name{ W.E.Pickett, S.E.Erwin, E.C.Ethridge}
(Preprint No. condens-matter/9611225).
%
\bibitem{Good1}
\Name{G.B.Goodenough} \Book{Magnetism and the Chemical bond} 
(Interscience Publishers, N.Y.) \Year{1963} \Page{269}. 
%
\bibitem{Good2}
\Name{G.B.Goodenough, G.Dutta, and A.Manthiram}
\Review{Phys. Rev. B.} \Vol{43} \Page{10170} \Year{1991}.
%
\end{thebibliography}
\end{document}